\documentclass[]{piparticle-final}
\usepackage{graphicx}
\usepackage{amsmath}

\usepackage{cite} 
\usepackage{epstopdf}

\begin{document}

\def\dy227{Dy$_2$Ti$_2$O$_7$}

\volume{7}               
\articlenumber{070009}   
\journalyear{2015}       
\editor{A. Vindigni}   
\reviewers{M. Perfetti, Dipartimento di Chimica, Universit\'a di Firenze, Italy}  
\received{17 May 2015}     
\accepted{12 June 2015}   
\runningauthor{S.\ A.\ Grigera \itshape{et al.}}  
\doi{070009}         

\title{An intermediate state between the kagome-ice and the fully
  polarized state in \dy227}

\author{S.\ A.\ Grigera,\cite{inst1,inst2}\thanks{E-mail: sag2@st-and.ac.uk}
~R.\ A.\ Borzi,\cite{inst3}
D.\ G.\ Slobinsky,\cite{inst1}\thanks{Now at: Departamento de Ingenier\'{\i}a Mec\'anica, Facultad Regional La Plata, Universidad Tecnol\'ogica Nacional, 1900 La Plata, Argentina.}
~A.\ S.\ Gibbs,\cite{inst1}\\
R.\ Higashinaka,\cite{inst4}
Y.\ Maeno,\cite{inst5}
T.\ S.\ Grigera\cite{inst3}}
\pipabstract{
\dy227 is at present the cleanest example of a spin-ice material.  Previous 
theoretical and experimental work on the first-order transition between the kagome-ice 
and the fully polarized state has been taken as a validation for the dipolar spin-ice model.
Here we investigate in further depth this phase transition using ac-susceptibility and 
dc-magnetization, and compare this results with Monte-Carlo simulations and previous 
magnetization and specific heat measurements. We find signatures of an intermediate
state between the kagome-ice and full polarization.  This signatures are absent in
current theoretical models used to describe spin-ice materials.
}

\maketitle

\blfootnote{
\begin{theaffiliation}{99}
   \institution{inst1} School of Physics and Astronomy, University of St
  Andrews, North Haugh, St Andrews KY16\ 9SS, UK
   \institution{inst2}Instituto de F\'{\i}sica de L\'{\i}quidos y Sistemas
  Biol\'ogicos, UNLP-CONICET, 1900 La Plata, Argentina
   \institution{inst3}Instituto de Investigaciones Fisicoqu\'\i{}micas
  Te\'oricas y Aplicadas UNLP-CONICET and Departamento de F\'\i{}sica,
  Facultad de Ciencias Exactas, Universidad Nacional de La Plata,
  1900 La Plata, Argentina
   \institution{inst4} Graduate School of Science, Tokyo Metropolitan University, Hachioji, Tokyo 192-0397, Japan.
   \institution{inst5}Department of Physics, Kyoto University, Kyoto 606-8502, Japan.

\end{theaffiliation}
}

\section{Introduction}

Spin-ice materials are deceptively simple in their constitution:
classical Ising spins with nearest-neighbour ferromagnetic
interactions forming a pyrochlore lattice.  This crystal structure
can be thought as an alternating stack of kagome and triangular
lattices along the [111] direction.  The spins sit at the vertices
of tetrahedra and can point either to their center or towards the
outside. The magnetic frustration can be seen at the level of a
single tetrahedron: the energy is minimized by having two spins
pointing inwards and two outwards. This is the \emph{ice rule},
which corresponds exactly to the Pauling rules for protons in water
ice;  like the latter, it also leads to zero-point entropy, a
characteristic signature of spin-ice systems \cite{harris97}.

We have chosen to work on \dy227 as the cleanest example of a
spin-ice material. Its ground state properties can be well described
by a model with only an effective nearest neighbour exchange
interaction $J_{\rm si}$ of $\approx 1.1\,$K \cite{science-rev01}.
Within this framework, when one applies an external magnetic field
$H$ in [111] below 1 K, the polarization of the system will happen in
two steps. First, the spins in the triangular lattice that lie
parallel to [111] will orient along the magnetic field, removing
\emph{part} of the residual entropy but with no change in the
configurational energy \cite{hiroi03,udagawa02}. When the magnetic
moment of this sublattice has saturated, the magnetization $M$
cannot be further increased without breaking the spin-ice rule.
This leads to a plateau as a function of field at $M = 3.33\, \mu_{\rm B}$/Dy-ion, characteristic of
the \emph{kagome ice} state. At higher fields, the spins in the
kagome lattice are finally fully polarized, leading to a sudden
but continuous increase in $M$ towards its saturation. This
behavior was predicted theoretically and found in
Monte Carlo simulations \cite{harris98,isakov04}. In spite of this, something
different happens in real spin-ice materials.

\begin{figure*}
\centerline{\includegraphics[width=2\columnwidth]{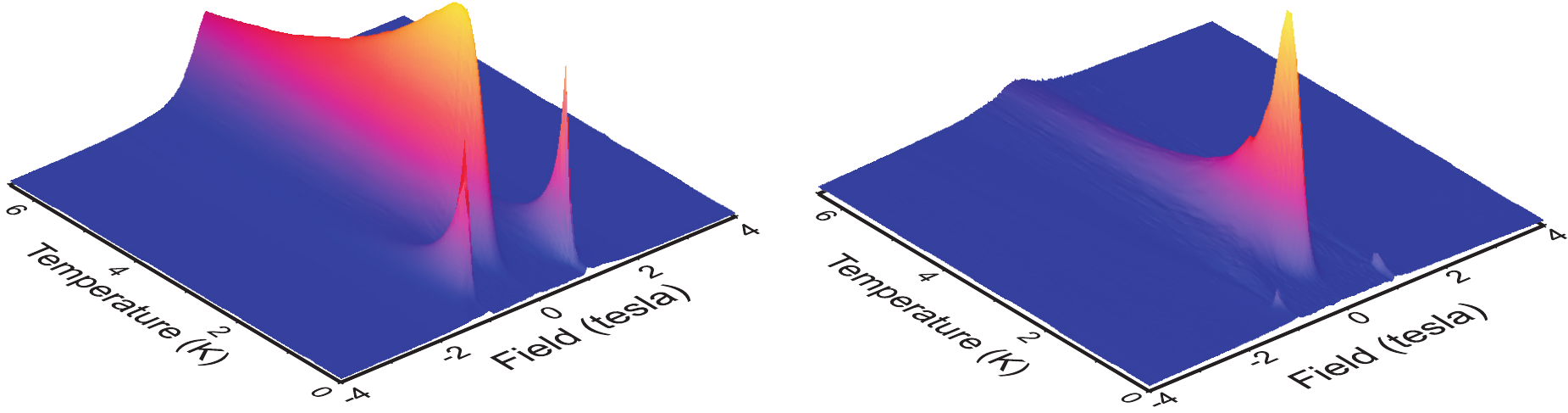}}
\caption{Real (left) and Imaginary (right) parts of the
  ac-susceptibility as a function of temperature and magnetic field,
  for temperatures between 50 mK and 6000 mK and magnetic fields
  between -4 and 4 T. The oscillatory field was of amplitude 0.05
  Oe and at a frequency of 87Hz. The zero-field Schottky-type anomaly corresponding to the
  onset of spin-ice correlations, and the peaks corresponding to the
  critical point at $\pm 1$ T and $\approx$400 mK are clearly seen.}
\label{3Dsusc}
\end{figure*}

Since the magnetic moment of the magnetic ions in spin-ice materials is
quite large -- the one associated with Dy$^{3+}$ ions in \dy227
is near 10$\mu_B$--, long range dipolar interactions have to be
considered \cite{bramwell01}. These interactions do not alter the zero field ground state \cite{isakov05}, but have a big effect on its excitations. In relation to this,
the transition to the fully polarized state ---which is
the main concern of this paper--- experiments a qualitative change.
T.~Sakakibara and collaborators \cite{sakakibara01} studied
experimentally the magnetization with $H
//$ [111] to temperatures much smaller than $J_{\rm si}$.  After a
well defined plateau at $\approx 3.33\, \mu_{\rm B}$/Dy-ion,
they observed a very sharp increase in the magnetization. The
presence of hysteresis was a convincing argument that the real
system reaches the fully polarized state through a metamagnetic
first order phase change at the lowest temperatures. The change in $M$
becomes continuous at the critical end-point $T_c = 360\,\pm 20$mK and
$\mu_0 H_c \approx 0.93\,$T \cite{sakakibara01}.

The change in character of this transition ---from a crossover to a
discontinuity when dipolar interactions are included--- was later
understood in terms of the defects associated with the breaking of
the ice rules, or \emph{monopoles}. The nearest neighbors model
corresponds to the case of free non-conserved monopoles sitting in
a diamond lattice. Including dipolar interactions implies
turning on a Coulomb interaction between these charges, allowing
them to condense through a real first order transition \cite{Cast_08}. Numerical simulations (including Ewald summations to
take into account long range interactions) proved this picture
right, and provided an additional validation to the dipolar model \cite{Cast_08}. The $M {\rm vs.\ } H$ curves obtained in these simulations
are quite symmetrical around $H_c$. The jump in
the magnetization $\Delta M(T)$ when crossing the first order
transition line grows very abruptly with decreasing temperature: for
$T$ only $\approx 10 \%$ below $T_c$, $\Delta M(T)$ amounts to
$\approx 90 \%$ of the total change in magnetization from the kagome ice to full saturation. In other words, almost full order is
achieved in the system for temperatures just below $T_c$ and a
magnetic field of 1 T.

Specific heat $C_p$ measurements confirmed the existence of a
critical end-point ---a sharp peak is clearly seen very near the
precise spot in field and temperature specified by Sakakibara {\em
et al.} \cite{higashinaka04}. However, the identification of a
single first-order line below $T_c$ is less clear.  The $C_p(T) vs.
H$ curves show peaks at the fields $H_c(T)$ identified in \cite{sakakibara01} as the first-order line, albeit of much smaller
amplitude than that at $T_c$.  Additionally, below 300 mK, a second
peak at higher fields is discernible \cite{higashinaka04}. Even at
the lowest temperatures ($100\,$mK), magnetic fields above 2 T are
needed to coerce the specific heat down to 0. This suggests that, in
spite of the absence of thermal excitations, the system does not
reach full polarization immediately after the first order transition from
the kagome ice, and an intermediate state establishes between these
two well-known phases. This specific heat features were confirmed by
$ac$-susceptibility measurements on the same samples \cite{thesis}.   
In all cases,  the sample sat at a fixed platform with respect to the 
magnetic field and therefore the alignment with respect to the [111]
direction was within a few degrees. An angular
dependent study of the magnetization with Sato and coworkers \cite{sato}
showed these asymmetries, and additional features in the polarization 
transition  were seen at small angles away from [111].   The implications
of these results in the current understanding and modeling of the spin-ice materials
have not been considered.

In this paper, we study in detail this additional intermediate state, and 
show that it cannot be explained by any of the models
 currently used to study spin-ice materials. Working at small 
angles away from  [111], we
looked for a magnetic signature by repeating the static
magnetization measurements in several samples. In addition,
improving the sensitivity by three orders of magnitude, we measured
ac-susceptibility at different frequencies, which also allowed us to 
do a characterisation of the dynamics of the observed transitions. 
In order to gain further insight into this possible intermediate state,
we performed Monte Carlo simulations of the experimental situation
using the currently accepted models including Ewald summations 
and exchange interactions up to the third
nearest neighbor \cite{Ruff05,Yaborskii08}.

\section{Methods}

For our work, we measured several \dy227 single crystals
grown in Kyoto and in St Andrews with the floating-zone method.
Samples were oriented using Laue diffraction and cut into $3\,$mm
long prisms of square or octagonal section of approximately
$1\,$mm$^2$, with the [111] direction along the long axis to reduce
demagnetising effects with the field in the vicinity of [111] ($\approx 5^o$).
 The experiments
were performed in a dilution refrigerator in St Andrews.  Samples
were thermally grounded to the mixing chamber through gold wires
attached into them with silver paint.  For susceptibility, we
used a drive field of $3.3\cdot10^{-5}\,$T r.m.s., and
counter-wound pickup coils each consisting of approximately 1000
turns of $12\,\mu$m diameter copper wire.  
The filling factor of the
 sample in the pick up coil was of approximately 90\%.
We measured using drive fields of frequencies varying from approx.~$10\,$Hz to $1.0\,$kHz.  
Low temperature transformers mounted on the $1\,$K pot
of the dilution refrigerator were used throughout to provide an
initial signal boost of approximately a factor of 100. The
magnetization was measured using a home-built capacitance
Faraday magnetometer \cite{magnetisation09}.

\section{Results and Discussion}

Figure~\ref{3Dsusc} shows the real ($\Delta\chi'$, left) and imaginary ($\Delta\chi''$, right)
parts of the ac-susceptibility $\chi$ as a function of temperature
and magnetic field in the whole area of interest.  The excitation
frequency in this case is $\omega = 87\,$Hz; the main features we
describe in the following are qualitatively independent of $\omega$.
At zero field, there is a very noticeable peak in both $\Delta\chi'$ and
$\Delta\chi''$ for $T \approx 2\,$K.  This corresponds to the
Schottky-type anomaly associated with the onset of spin-ice
correlations of the system. The magnetic field axis spans from -4 to 4
T, and we can clearly see in the real part two peaks 
(at positive and negative fields) corresponding to the
critical point at $\approx \pm 1\,$T and $\approx 400\,$mK. For
temperatures below $400\,$mK, we see a much smaller feature in $\Delta\chi''$, 
which has a correspondence in $\Delta\chi''$: a ridge
with an amplitude that decreases as a function of temperature. The
magnitude of the latter is comparatively very small. At low
temperatures and for fields $0.3\,$T$ < |\mu_0 H| <
0.9\,$T, and $2\,$T $< |\mu_0 H|$, the
susceptibility is very low, in accordance to the kagome ice plateau and
the saturation in the magnetization, respectively.

\begin{figure}
\centerline{\includegraphics[width=\columnwidth]{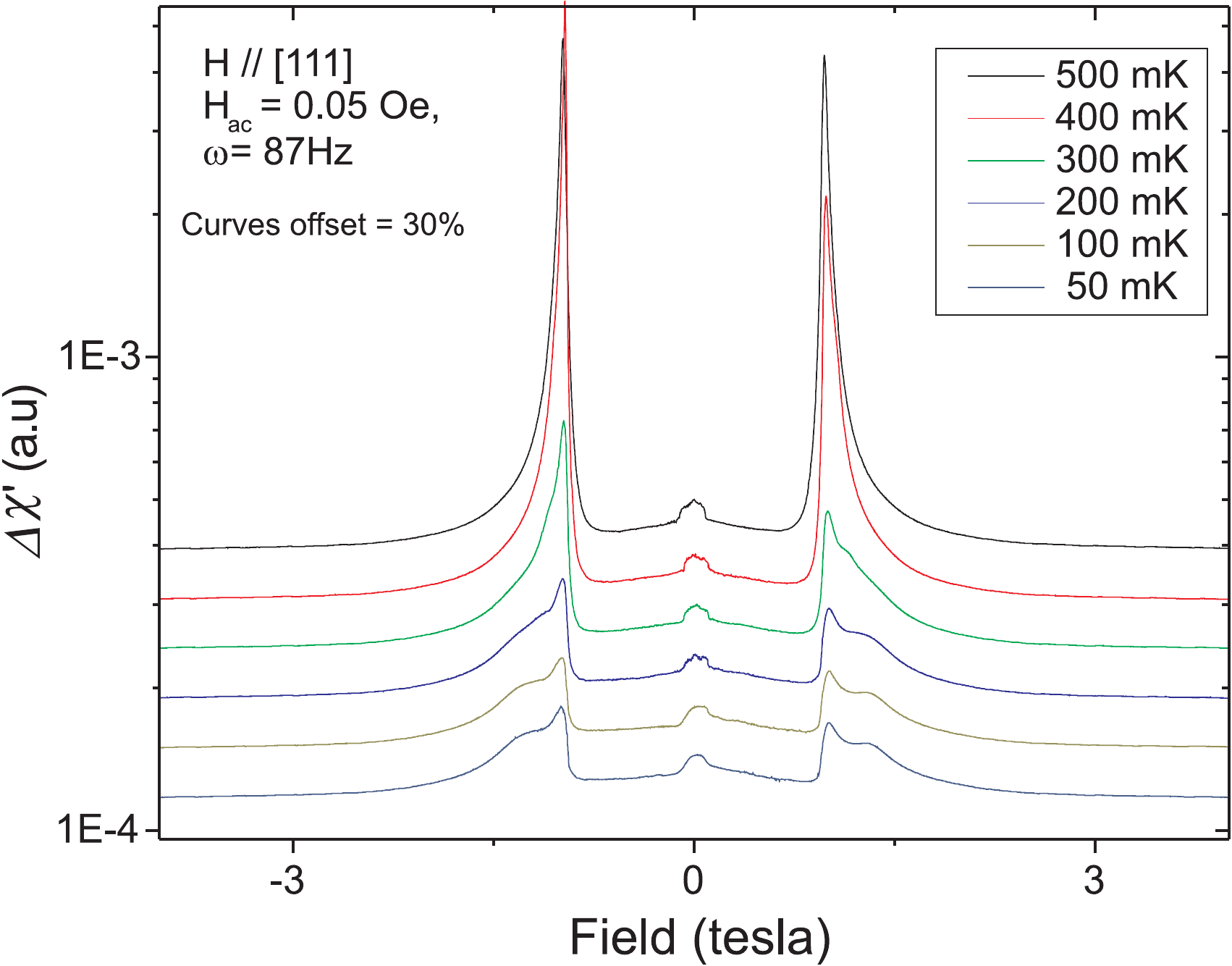}}
\caption{Low temperature real part of the ac-susceptibility as a
  function of field at fixed temperatures as indicated in the plot.
  The excitation field was 0.05 Oe at a frequency of 87Hz. The curves
  are offset by 30\% for clarity.  As temperature is lowered from 400
  mK, the peak at approx. 1 T at 400mK rapidly decreases in
  amplitude, and splits into two peaks at lower
  temperatures.}
\label{lt-susc}
\end{figure}

We now concentrate on the real part of the susceptibility at temperatures 
below $J_{\rm si}$.  In Fig. \ref{lt-susc}, we can see a series of curves at fixed temperatures
(from 50 to $500\,$mK) and fields between $-3.5$ and $3.5\,$T. The
excitation field used was $0.05\,$Oe, and the frequency $87\,$Hz.  The
curves have been offset by 30\% for clarity.  The field was swept from
negative to positive values.  Before the kagome ice state is
established, the low field susceptibility ($|\mu_0 H| < 0.3\,$T) at
temperatures below $600\,$mK is strongly dependent on the magnetic
field sweep rate and direction (increasing or decreasing), both signs
of out-of-equilibrium behavior. At higher magnetic fields, we only
observe a small difference in the height of the peaks at around $\pm 1\,$T,
depending on whether the transitions are swept upwards or downwards in
field. The position changes very little, and the shape of the features
is unaltered.  As we lower the temperature, the peak at $\approx 1\,$T decreases markedly in amplitude, but without
a corresponding change in its high field side shoulder. Below $400\,$mK, it eventually splits into two
distinct features. Their separation in the field axis ($\approx$ 0.1 T at 300 mK) is consistent
with previous measurements for a similar sample orientation with respect to [111] \cite{sato}. 
While the first set of peaks has a correlate in the imaginary part of
$\Delta\chi$ (not shown here), no feature is discernible in $\Delta\chi''$ for the
peaks at higher fields.

\begin{figure}
\centerline{\includegraphics[width=\columnwidth]{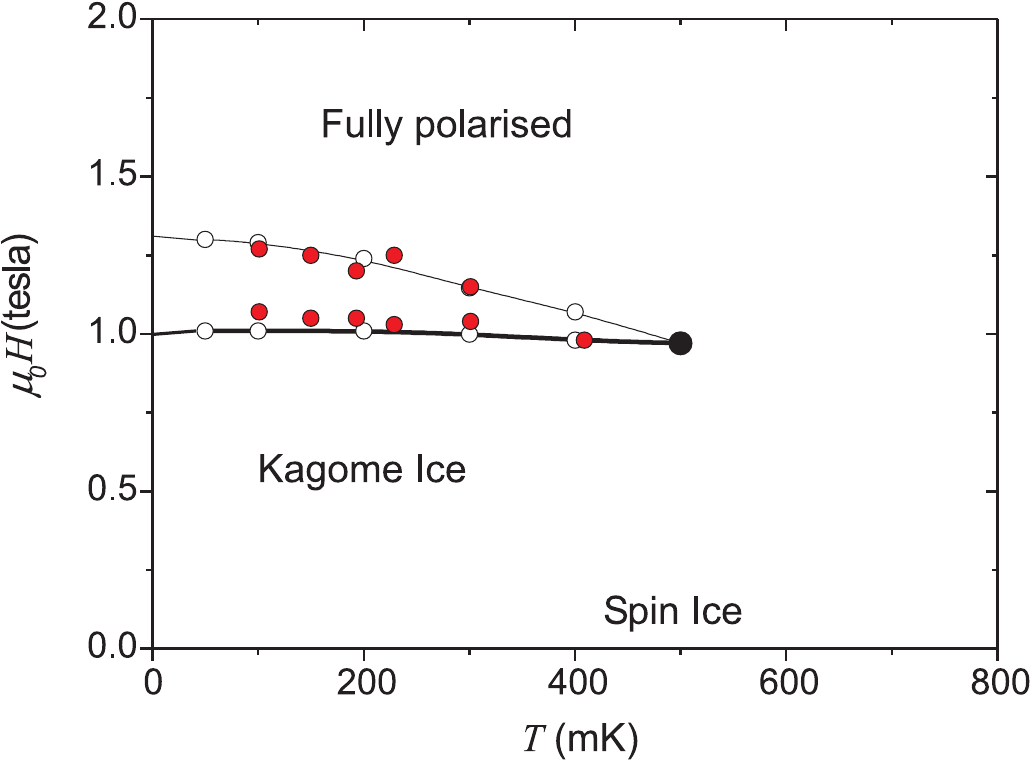}}
\caption{Phase diagram with field  slightly tilted from  [111] ($\theta \leq 5^o$).
  An intermediate phase
  is seen between the kagome-ice and fully polarized regions.  The
  black circle is the critical point as identified from a peak in the
  real part of the ac-susceptibility, $\chi'$.  The dotted white
  circles correspond to a small doubled peaks seen in $\chi'$ with a
  corresponding feature in the imaginary part $\chi''$, while the
  white circles denote small peak in $\chi'$ with no signature in
  $\chi''$.  The red circles are taken from peaks in the specific heat
  ($C$) measurements of reference \cite{higashinaka04}.  The
  main divergence of $C$ seen in reference \cite{higashinaka04}
  and identified as a critical point coincides with the critical point (black
  circle).}
\label{ph-d}
\end{figure}

In Fig. \ref{ph-d}, we have plotted the position of these peaks as a
function of field and temperature (white circles), and the position of
the critical point (black circle).  We have taken the specific heat
data from reference~\cite{higashinaka04} and determined the
position of the peaks in $C$ {\sl vs.\/} $H$ for different
temperatures.  These are plotted in this same graphic as red symbols.
The coincidence between these two experiments of different quantities,
on different samples, laboratories and experimental setup is 
remarkable.

\begin{figure}
\centerline{\includegraphics[width=\columnwidth]{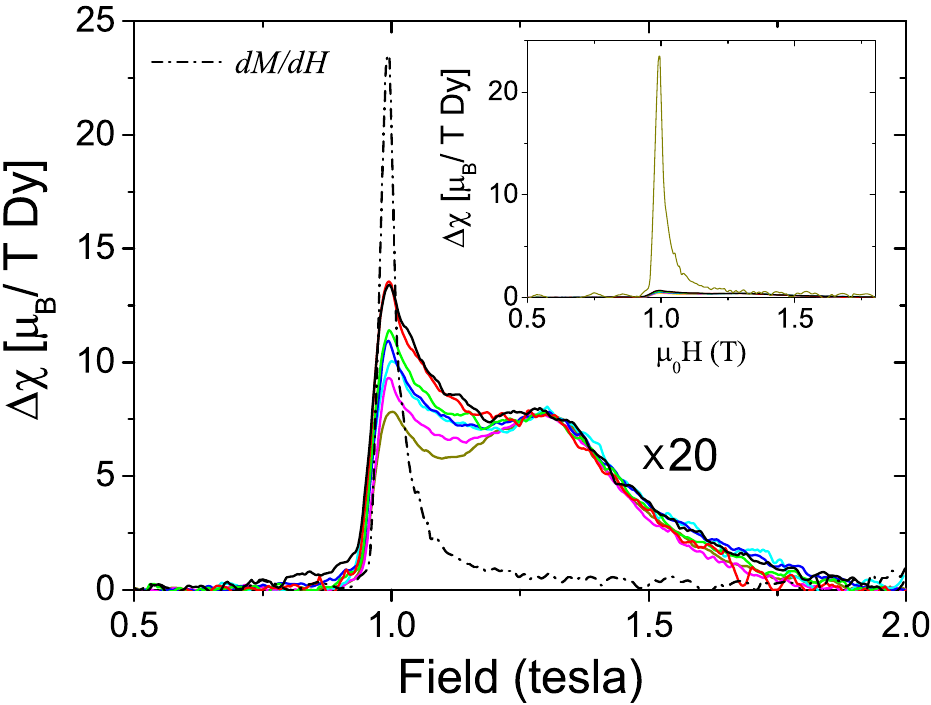}}
\caption{$dM/dH$ (dotted line) and real part of ac-susceptibility
measured at different excitation frequencies, from top to bottom:
19, 37, 77, 136, 277, 561, and 1117 Hz, and for $T = 100 {\rm mK}$.
For ease of comparison, the latter have been multiplied by a factor
20, and normalized to the amplitude of second peak (no imaginary
part has been measured for this feature). The inset shows both sets
of data in the same scale.} \label{freq}
\end{figure}

As mentioned before, this secondary peak at higher fields is absent in the $dM/dH$ data presented on Ref. \cite{sakakibara01}.
We measured the magnetization using a Faraday balance on the same samples and under similar temperature and field
conditions than before \cite{shift}. The main body of Fig. \cite{ac-loss} shows our 
$dM/dH$ as a function of field, compared with curves
of $\Delta\chi$ at $T = 100\,$mK and frequencies spanning two orders of magnitude
(from $\omega \approx 10$ to $1000\,$Hz). For clarity, we have multiplied $\Delta\chi$ by a factor of twenty.
The peak in $dM/dH$ is markedly asymmetric, with an extended tail in the high field
side but no additional feature is seen at high fields, in coincidence
 with Sakakibara's observations. On the other hand, the
second peak is clearly seen for low
temperature ($T < 400\,$mK) at all measured frequencies in the
ac-susceptibility. While these two experiments seem to be in mutual contradiction, 
the issue can be easily explained in terms of  the resolutions of both techniques.  
The inset of Fig. \ref{freq} shows both sets of data on the same scale; 
we can see that the high field shoulder on the $dM/dH$ peak directly corresponds 
(in the limit of long measurement times or low frequencies) to the second peak detected with ac-susceptibility.

Through this analysis, we can see that the experimental volume of data
concerning this transition seems to be compatible. Between $\approx
300 mK$ and the lowest temperatures (50 mK in Ref. \cite{sakakibara01}),
only $\approx 60\%$ of the total change in magnetization occurs
when traversing the first order transition line. The remaining 40\%
is delivered \emph{gradually} when the field is further increased to
values well above 1.5$H_c$, in a fashion that does not seem to
depend much on temperature (see Fig. 3 on Ref. \cite{sakakibara01}). 
This gradual (as opposed to
discontinuous) change is behind the asymmetric shape of the
magnetization curves, and the second set of peaks in $C_p$ and
$\Delta\chi$.

The theoretical prediction for the transition between kagome-ice to
fully polarized state with field in [111] was of a single
transition---the ``dimer to monomer'' transition of
refs.~\cite{moessner03,isakov04}.  A small additional perpendicular field
-- present in the experiments at small angles away from [111]-- induces 
order in the dimers in the kagome-ice state, but does not
change the prediction of a single transition into the fully 
polarised ``monomer'' state \cite{moessner03}.
This might hold true when further interactions are added, such as
dipolar or further neighbor exchange interactions.  In order to investigate
this, we performed a numerical check. We did extensive Monte Carlo simulations of the dipolar
model including Ewald summations to account for the dipolar long
range interactions. We also added exchange interactions up to third
nearest neighbors (taking the exchange constants and other
parameters within the constraints given by refs. ~\cite{Ruff05,Yaborskii08}). We explored
a wide range of field angles around [111], but were unable to detect
a double feature in $C_V$ at low temperatures compatible with the
experimental observations. It is then worth stressing that the very observation
of a second feature --even when taking into account a possible sample misalignment-- 
asks for new ingredients in the Hamiltonians that are regularly used to describe 
spin-ice materials.

Given these considerations, it is difficult to discuss on the nature of this
intermediate state. It is tempting to think of 
some sort of ``charge'' ordering in the diamond lattice (2-in 2-out tetrahedra
within a majority of 3-1 and 1-3), previous to the final Zn-blende arrangement, where only $\approx 40-50 \%$ of 
the sites are occupied by single monopoles.  Note
that this does not rule out the possibility of still storing some
residual entropy, since there are different spin configurations that generate the
same charge within a given tetrahedron. We have not found previous data of the evolution of
the entropy as a function of field at temperatures well below $T_c$.
However, the very asymmetric shape of the entropy at 350 mK obtained
using the magnetocaloric effect shows that at this temperature the
system is already experiencing a strong first order metamagnetic
transition, as mentioned in Ref. \cite{magnetoc}. This work shows that a big
fraction of the residual entropy stored in the kagome planes remains
in the system well above $H_c$ \cite{magnetoc}, suggesting that the
intermediate state is indeed a partially disordered one.

\section{Conclusions}

In conclusion, we observe an intermediate state between the kagome-ice
and the fully polarized state when the field is slightly tilted from the [111] direction. The signature
of a double step we find in ac-susceptibility and magnetization measurements is also
present in earlier calorimetric measurements, and suggested by magnetocaloric
effect experiments. This feature cannot be captured by the models
regularly used to describe spin-ice systems, fact that asks for further model refinements. 
At present, this data stands as a challenge for the development 
of a realistic theoretical model of spin-ice materials.

\begin{acknowledgements}
We thank Joseph Betouras, Andrew Green and  Chris Hooley for useful
discussions.  SAG would like to acknowledge
financial support from the Royal Society (UK), RAB and TSG
from CONICET, UNLP and ANPCYT (Argentina).
\end{acknowledgements}


\begin{thebibliography}{}
\bibitem{shift} Since we are using two different probes, the exact position
and orientation of the sample with respect to the magnet differs slightly 
between the ac-susceptibility and the magnetization measurement, and a positive 
shift of 0.02 T in the magnetization measurement was necessary to make 
the critical field of the first order transition coincide.
\bibitem{ac-loss} The jump in magnetization $\Delta M$ is
essentially independent of temperature at low $T$. Since $\Delta M$
is the integral of the susceptibility, one would naively expect that
area below the peak in $\chi'$ (Fig.~\ref{lt-susc}) to be also
independent of $T$. But this is true only for the dc susceptibility,
or, more accurately, for $\chi'$ measured at frequencies lower than
the inverse of the longest relaxation time. The fact that we can
measure an out of phase response $\Delta \chi''$ reveals we are
actually measuring dynamic response, i.e.\ that our frequencies are
high and some relaxation processes do not contribute to
$\Delta\chi'$. \raggedbottom 
\pagebreak Since relaxation times grow on lowering the
temperature, the area loss observed in this figures is quite
natural.
\end{thebibliography}
\end{document}